# High pressure growth and electron transport properties of superconducting SmFeAsO$_{1-x}$H$_x$ single crystals

Soshi Iimura[1, *], Takashi Muramoto[1], Satoru Fujitsu[2], Satoru Matsuishi[2], and Hideo Hosono[1, 2, *]

[1]Laboratory for Materials and Structures, Institute for Innovative Research, Tokyo Institute of Technology, Yokohama 226-8503, Japan

[2] Materials Research Center for Element Strategy, Tokyo Institute of Technology, Yokohama 226-8503, Japan

Correspondence and requests for materials should be addressed to S. Iimura and H. Hosono.

S. Iimura, e-mail: s_iimura@lucid.msl.titech.ac.jp

H. Hosono, e-mail: hosono@msl.titech.ac.jp

**Abstract**

We report the single crystal growth and characterization of the highest $T_c$ iron-based superconductor SmFeAsO$_{1-x}$H$_x$. Some sub-millimeter-sized crystals were grown using the mixture flux of Na$_3$As + 3NaH + As at 3.0 GPa and 1473 K. The chemical composition analyses confirmed 10% substitution of hydrogen for the oxygen site ($x$ = 0.10), however, the structural analyses suggested that the obtained crystal forms a multi-domain structure. By using the FIB technique we fabricated the single domain SmFeAsO$_{0.9}$H$_{0.10}$ crystal with the $T_c$ of 42 K, and revealed the metallic conduction in in-plane ($\rho^{ab}$), while semiconducting in the out-of-plane ($\rho^c$). From the in-plane Hall coefficient measurements, we confirmed that the dominant carrier of SmFeAsO$_{0.9}$H$_{0.10}$ crystal is an electron, and the hydride ion occupied at the site of the oxygen ion effectively supplies a carrier electron per iron following the equation: O$^{2-}$ → H$^-$ + $e^-$.

**Introduction**

Since the discovery of high critical temperature ($T_c$) superconductivity at 26 K in LaFeAsO$_{1-x}$F$_x$ at the end of February 2008, iron-based superconductors have attracted widespread research interests across fields of engineering and basic science, due to their promising characteristics for practical applications and puzzling electronic properties [1–3]. Although a variety of the parent materials have been reported so far, the crystal structure shares a common conducting [Fe*Pn*]$^-$ (*Pn* = pnictogen) or [Fe*Ch*]$^-$ (*Ch* = chalcogen) layer composed of the edge-sharing Fe*Pn*$_4$ or Fe*Ch*$_4$ tetrahedra. Particularly, the iron pnictides family consists of various structures by inserting alkali metal cation, alkaline earth metal cation, positively charged PbO-type layer, or perovskite-type block between the Fe*Pn* layers to compensate the neutrality. A prototypical *RE*FeAsO-type (*RE* = rare earth) material, abbreviated as 1111-type, crystallizes in a tetragonal lattice at room temperature (space group *P*4/*nmm*, $a \sim 4$ Å, $c \sim 8$-9 Å, and $Z = 2$), in which the conducting [FeAs]$^-$ and insulating [*RE*O]$^+$ layers stack alternately along the crystallographic *c* axis (**Fig. 1a**). On cooling, they undergo a tetragonal-orthorhombic transition at $T \sim 150$ K, accompanying a paramagnetic-antiferromagnetic (PM-AFM) transition just below the structural transition temperature [4,5]. Superconductivity appears if the AFM is suppressed by doping carriers or applying pressure (see $x = 0$ in **Fig. 1b**).

Carrier doping mode in 1111-type is categorized into four from a view of doping site and carrier polarity. Substitution for the site in conducting layer is called "direct doping", while that in insulating layer is called "indirect doping". Among those methods, the indirect electron doping has intensively been studied because it induces higher $T_c$ than the other methods due to less influence of structural randomness on the conduction plane, and the optimum $T_c$ exceeding 50 K in late *RE*-substituted 1111-type records the highest value in each of iron-based superconductors [6,7].

At the early stage, fluoride ion substation for the oxygen site ($O^{2-} \rightarrow F^- + e^-$) was the standard way of the indirect electron-doping inducing superconductivity in the 1111-type compounds [1,8]. However, the precipitation of stable *RE*OF prohibits further electron doping and exploration of their physical properties over $x = 0.2$ [9]. In 2010, we proposed a hydride-ion-substitution-method ($O^{2-} \rightarrow H^- + e^-$) as an alternative way of the electron doping [10]. This method expanded the doping limitation to $x \sim 0.5$, and made possible to access the overdoped region where the superconductivity disappears by excess carrier doping. Systematic chemical composition analyses on both the oxygen and hydrogen, and a neutron diffraction measurement on the deuterium substituted sample demonstrate the substitution of hydrogen for the oxygen site [10,11]. Moreover, density functional theory calculations and the Hall coefficient measurement confirm that the hydrogen in the oxygen site behaves as an anion, and supply an electron to the Fermi level composed mainly of Fe-3$d$ orbitals [11,12]. Currently, therefore, this method serves an only way to enable to investigate the electronic phase diagram of 1111-type, covering the AFM phase at $x \sim 0$ (AFM1), the superconducting phase in the $x$ ranging $0.05 \leq x \leq 0.45$ (SC), and the overdoped regime with $x > 0.45$. It was our surprise to see a structural transition from the tetragonal to a unique orthorhombic structure in AFM2 phase(**Fig. 1b**) [13–15].

Despite of these recent progress in the synthetic methods and the attracting physical properties, the single crystal growth of 1111-type compounds has been a difficult task because of a lack of effective flux and their severe synthesis condition requiring high temperature. Karpinski's group have adopted a high temperature and high pressure technique and succeeded in the growth of fluorine substituted 1111-type single crystals using NaCl/KCl as a flux [16–18]. This method has several advantages in comparison with the conventional ampoule method, since it avoids vaporization losses and allows control of the composition even at the high temperatures required for single-crystal growth. However, even at optimal conditions, the growth rate and the doping level are extremely low. Furthermore bipolar nature of hydrogen in $RE$FeAsO$_{1-x}$H$_x$ raises the difficulty [19]. In order to clarify the nature of highest $T_c$ superconductivity in $RE$FeAsO$_{1-x}$H$_x$, it is imperative to establish a new growth condition of the 1111-type and investigate the physical properties using the single crystal.

Here, we report the flux growth and the characterization of crystal structure and physical properties of single crystals of the highest $T_c$ superconductor SmFeAsO$_{1-x}$H$_x$. We undertake factors controlling the crystal growth, including the kinds of flux, synthesis pressure and starting chemical composition, and report the results of resistivity, susceptibility, and Hall coefficient measurements on the single domain crystal fabricated by the focused ion beam (FIB) technique.

**Experimental**

**Single crystal growth**

Single crystals of SmFeAsO$_{1-x}$H$_x$ were grown using a mixture of precursors, Sm$_2$O$_3$, SmH$_2$, SmAs, Fe$_2$As, and FeAs, according to the equation [10]:

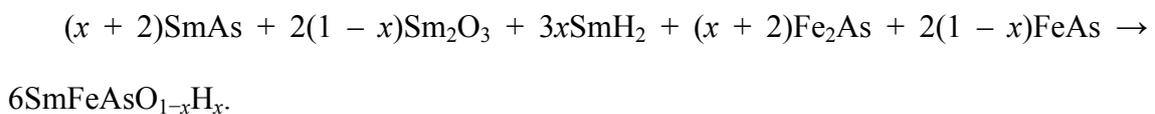

$(x + 2)$SmAs $+ 2(1 - x)$Sm$_2$O$_3$ $+ 3x$SmH$_2$ $+ (x + 2)$Fe$_2$As $+ 2(1 - x)$FeAs $\rightarrow$ 6SmFeAsO$_{1-x}$H$_x$.

The precursors were grounded with Na$_3$As, 3NaH + As, or their mixture flux in a dehydrated silica glass agate mortar for 1 hour inside an Ar-filled glovebox. The mixture was precompressed with a load of 4 MPa to form a pellet (6 mm in diameter and 2.5 mm in thickness) and afterward loaded into the sample units for the high pressure synthesis. **Figure 1c** and **1d** provide the schematic illustration of the sample cell assembly for belt-type high pressure apparatus used in this study. The sample cell is mainly composed of 90wt%-NaCl and 10wt%-ZrO$_2$. The graphite sleeve is used as a resistance heater touching two Mo discs for electric lead below and above the cell. The boron nitride (BN) sleeve of 6 mm (7 mm) in inner (outer) diameter and 8 mm in length is inserted to the NaCl (+10wt%ZrO$_2$) tube, and the sample pellet is placed inside the BN crucible. For the single crystal growth using the Na$_3$As flux and the synthesis of polycrystalline samples, two pellet (6 mm in diameter and 2 mm in thickness) of mixture of NaBH$_4$ and Ca(OH)$_2$ with a molar ratio of 1 : 2 was placed below and above the sample pellet (**Fig.1c**) to supply excess hydrogen through the following reaction [20,21]:

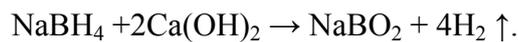
$$\mathrm{NaBH_4 + 2Ca(OH)_2 \rightarrow NaBO_2 + 4H_2 \uparrow}.$$

For the crystal growth using the 3NaH + As or Na$_3$As + 3NaH + As flux, we adopted a more simple assembly shown in **Fig.1d**, since we assumed that the flux itself releases the excess hydrogen through the following reaction:

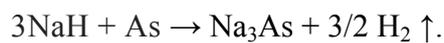
$$\mathrm{3NaH + As \rightarrow Na_3As + 3/2\, H_2 \uparrow}.$$

The temperature at the sample position was estimated by the predetermined relation between applied electrical power and temperature measured by Pt-PtRh thermocouples at 2.5 and 5.5 GPa, while the pressure calibration was conducted against the well-known phase transitions of Bi (I-II) at 2.55, Bi (II-III) at 2.69, and Ba (II-III) at 5.5 GPa at room temperature [22]. Recovered samples were immersed into deionized and UV-irradiated water (Milli-Q Integral3, Merck Millipore) to dissolve the flux.

**Chemical composition and structural analyses**

The chemical compositions of the SmFeAsO$_{1-x}$H$_x$, except for hydrogen, were measured by an electron probe micro-analyzer (EPMA, JEOL JXA-8530F), while the hydrogen content was determined by the thermal gas desorption spectrometry (TDS, ESCO TDS-1400TV). The quantitative analyses using EPMA were conducted on the single crystal using the atomic number, absorption, and fluorescence (ZAF) correction method with the standard samples of SmP$_5$O$_{14}$ for Sm, elemental Fe for Fe, LaAs for As, and Gd$_3$Ga$_5$O$_{12}$ for O. The TDS were performed on the grounded crystals picked up from each batch. For the estimation of hydrogen contents, we used a standard sample that implants $1.06 \times 10^{17}$ hydrogen ion into silicon wafer. The crystalline phase and the quality of the samples were determined by X-ray diffraction (XRD, Bruker AXS, D8 ADVANCE-TXS, source: CuKα radiated from a rotational anode, θ-2θ scan Bragg-Brentano geometry). The fluctuations of the crystallite orientations were characterized using the rocking curves of out-of-plane (2θ-fized ω scans) diffractions with a high-resolution XRD apparatus (RIGAKU, SmartLab, source: CuKα$_1$ monochromated by Ge(220)). The in-plane orientation of the films was investigated using a pole figure geometry by the high-resolution XRD apparatus.

**Physical property measurements**

Transport measurements were performed using a physical properties measurement system (PPMS, Quantum Design, Inc.). The electrical resistivity, ρ, and Hall effect were measured using a four-probe and a five-probe techniques, respectively, in which the fifth voltage lead for the Hall measurement is attached in parallel to one of the other voltage leads to null an artificial voltage depending on the sample resistance and/or magnitude of the electric bias field. Conducting leads were deposited by FIB (JEOL JEM-9320FIB); $W(CO)_6$ gas was injected from the needle and adsorbs onto the sample surface. The $Ga^+$ beam decomposes the gas, which leaves a deposited layer of W, while the CO is removed through the vacuum system. Zero field cooling susceptibility was measured on the grounded crystals under the magnetic field (H) of 10 Oe using a magnetic properties measurement system (MPMS, Quantum Design, Inc.).

**Results & Discussion**

A) **Optimization of growth condition of SmFeAsO$_{1-x}$H$_x$ single crystals**

**Table 1** summarizes the growth conditions (GC) applied in this study. To prevent loss of hydrogen during synthesis, we grew all of crystals for a quite short time (30 min) without temperature gradient. Previous researches on single crystal growth of the 1111-type propose NaCl, KI, KCl, NaAs, and KAs as an effective flux [16-19], however, we couldn't obtain over 10μm-sized crystals using the alkali halides fluxes for the short time, while the NaAs seemed to impede the formation of 1111-type phase probably because it is reactive with the SmH$_2$ precursor. For these reasons, we choose Na$_3$As, 3NaH + As, and those mixture as the flux, and optimized other parameters. **Figure 2(a) and (b)** show the optical microscope images of SmFeAsO$_{1-x}$H$_x$ single crystal grown by GC1. Some submillimeter-sized crystals were found even for the short-time heating, indicating that the Na$_3$As is effective to grow the 1111-type crystals, and superior to those alkali halides and the alkali monoarsenides fluxes in that the large size crystals are obtained more rapidly. **Figure 2(c) and (d)** show the images of crystal grown by GC2 using the 3NaH + As flux. Unlike the GC1, the average size of crystal was decreased, and more prominently, a mosaic structure due to a multi-nucleation was observed in the larger crystal (**Fig. 2(d)**). From the results of composition analyses on the crystals grown by GC1 and GC2, we found that the inclusion of NaH increases the hydrogen contents, though the amount of oxygen vacancy is unchanged compared with the case of Na$_3$As. Based on these results, we concluded that the Na$_3$As flux is more efficient in obtaining bigger crystals, while it makes hydrogen doping more difficult. Thus, we tested the mixture flux (Na$_3$As + 3NaH + As) to grow a higher-doped and larger-sized crystal (GC4). In **Fig. 2f**, an over 100 μm size crystal is shown, and the results of chemical composition analyses are consistent with the chemical formula of SmFeAsO$_{1-x}$H$_x$ with $x = 0.10$.

Increasing the growth pressure from 3.0 GPa to 5.5 GPa promotes the hydrogen substitution to $x = 0.16$ for the condition using 3NaH + As flux (GC5) and 0.18 using the mixture flux (GC6), whereas it seems to interfere the growth of larger crystals with the lateral size over 100 μm (**Fig. 2g and 2h**). In order to obtain much higher doped crystal, the nominal hydrogen content was increased as following the chemical equation; $(x + 2 - 3xy)$SmAs + $2(1 - x)$Sm$_2$O$_3$ + $3(x + xy)$SmH$_2$ + $(x + 2 - 3xy)$Fe$_2$As +

$2(1 - x + 3xy)$FeAs → $6$SmFeAsO$_{1-x}$H$_{x(1 + y)}$, where the $y$ denotes the excess hydrogen contents (GC7-9). However, the actual hydrogen content in those crystals could not exceed the value of 0.18 obtained by the GC6.

**B) Structural analysis**

Next, we characterized the structure of the ~ 100 μm crystal obtained by GC4 by the XRD measurements. The out-of-plane θ-2θ scan clearly detected the 00$l$ series diffractions ($l$ = 3, 4, 5, and 6) (**Fig. 3a**), whereas some asymmetric, broad peaks were observed in the rocking curve of 003 reflection (**Fig. 3b**). The 012 pole figure shows a four-fold symmetric diffraction due to the tetragonal symmetry of SmFeAsO$_{1-x}$H$_x$ every φ = 90 deg., as well as several weak diffractions between the strong 012 diffractions (**Fig. 3c and 2d**). These results suggest that the obtained crystal forms a multi-domain structure. The *c*-axis length was estimated by Pawley method from the θ-2θ scan profile to be *c* = 8.456(2)Å, while the *a* and *c*-axis lengths estimated from the grounded crystals picked up from the GC4-batch were 3.9248(2) and 8.465(3) Å, respectively, in good agreement with those of polycrystalline samples at $x$ = 0.13 shown in **Fig. 3e and 3f** [10].

**C) Transport and susceptibility measurements**

Since the grain boundary in crystal scatters the conducting electrons and inhibits to see the intrinsic transport properties, we employed the FIB technique to fabricate a single domain sample for transport measurements [23]. In **Figure 4(a)**, we show the secondary electron microscope (SEM) image of resistance bar used for the anisotropic resistivity measurements. First, we cut the single crystal using the FIB into a smaller bar with ~20 μm in the *ab* plane and 10μm along with the *c* axis. Then, it was glued on the Au-electrodes printed on a GaAs substrate using an epoxy adhesive for low temperature application, and then, the tungsten electrodes were deposited on both the edge of bar and the Au-electrodes to make the electric contact. Finally, we obtained the U-shaped sample by cutting the needless parts. The distance and the cross-section between the tungsten voltage electrodes for the out-of-plane resistivity ($\rho^c$) measurement (V($c$2)) is 3.33 μm and 4.76 μm², respectively, while those for the in-plane ($\rho^{ab}$) (V($ab$)) is 8.95 μm and the cross-section is 11.19 μm², respectively. **Figure 4(b)** shows the resistance bar used for the five-probe Hall measurement in the *ab* plane. The voltage lead separation is 15.8μm, while the thickness is 2.3μm.

Temperature dependences of the $\rho^{ab}$ and $\rho^c$ resistivity are shown in **Fig. 4(c)**. The $\rho^{ab}$ at $T$ = 300 K is less than one-tenth of the resistivity of polycrystalline SmFeAsO$_{0.87}$H$_{0.13}$, and three times smaller than the $\rho^c$. The metallic conduction with a convex curvature of $\rho^{ab}$ ($T$) is similar to the polycrystalline case, while in contrast, the $\rho^c$ ($T$) behaves semiconducting. The resistivity anisotropy ($\rho^c/\rho^{ab}$), reflecting the crystal and electronic structures of material, reaches 7.8 at $T$ = 50 K. In Table 2 we summarize the $\rho^c/\rho^{ab}$ at $T$ = 50 K of several iron-based superconductors [23–31]. The slightly smaller $\rho^c/\rho^{ab}$ in SmFeAsO$_{0.9}$H$_{0.10}$ ($\rho^c/\rho^{ab}$ = 7.8) than SmFeAsO$_{0.7}$F$_{0.25}$ ($\rho^c/\rho^{ab}$ = 8.4) may reflect an increased three dimensionality of the electronic structure in hydrogen-substituent proposed theoretically in CaFeAsH and CaFeAsF [32]. For other materials, the anisotropy increases as the inter-layer separation increases. The exceptionally large values of K$_{0.58}$Rb$_{0.42}$Fe$_{1.72}$Se$_2$ and (Li$_{0.84}$Fe$_{0.16}$)OHFe$_{0.98}$Se may reflect the effect of a magnetic scattering and an electronic transition above 50 K, respectively [29,30]. The superconductivity appears at 42 K, and the zero resistivity is attained at 41 K (**inset of**

**Fig.4(c)**). A strong diamagnetic response due to shielding effect was observed below $T =$ 43 K, consistent with the $T_c$ measured by the resistivity measurement (**Fig.4(d)**). The too large volume fraction exceeding 100% is due to the large demagnetization over the surface of the plate-like sample [33].

We analyzed the temperature dependence of $\rho^{ab}$ by fitting to a power law function: $\rho^{ab} = \rho^{ab}_0 + AT^n$. The transport properties above $T_c$ of high-temperature superconductors have been intensively studied since the discovery of copper oxide superconductors [34,35]. It is well known that in the overdoped state of high-$T_c$ superconductors, the temperature dependence of resistivity follows Fermi liquid behavior with a quadratic dependence of $n = 2$ in low temperature. Meanwhile, the linear $T$-dependence ($n = 1$), referred to as non-Fermi liquid state, is frequently observed in the normal conducting state of superconducting samples that especially show an optimum $T_c$, even though the nature of non-Fermi liquid state is in debate. Therefore, a variation from the $T$- to $T^2$-dependece with doping is considered as one of common characteristics in the cuprates and iron-based superconductors. On the other hand, the square-root relationship ($n = 0.5$) is not established well, and recently proposed theoretically to be a signature of a spin-frozen regime where the fluctuations of spin are very slow (the spin susceptibility has Curie-Weiss form and a large static value) while those of orbital are very fast (the orbital susceptibility is small and is less dependent on temperature change) [36]. This behavior is pronounced at the special valence of one unit of charge away from half filling, *i. e.*, $d^6$ state of $Fe^{2+}$ and $Ru^{2+}$, observed experimentally in very few compounds, such as $FeTe_{1-x}Se_x$, polycrystalline $LaFeAsO_{0.6}H_{0.4}$, and $Sr_2RuO_4$ [12,37–39]. The convex curvature of $\rho^{ab}$ ($T$) with the $n = 0.47$ suggest that the normal conducting state of SmFeAsO$_{0.9}$H$_{0.10}$, which corresponds the boundary between the AFM1 and SC phases, may be located in the spin-frozen state.

The Hall coefficient, $R_H$, of SmFeAsO$_{0.90}$H$_{0.10}$ shows a negative sign above the $T_c$, indicating that the dominant carrier in the normal conducting state is electron (**Fig.(f)**). If we apply a two-band model to the semi-metallic band structure of SmFeAsO$_{1-x}$H$_x$, the $R_H$ is expressed as [40]

$$R_H = \frac{1}{e}\frac{(n_e\mu_e^2 - n_h\mu_h^2)n_e\mu_e^2 + \mu_e^2\mu_h^2(n_e - n_h)(\mu_0 H)^2}{(n_e\mu_e + n_h\mu_h)^2 + \mu_e^2\mu_h^2(n_e - n_h)^2(\mu_0 H)^2} \quad (1)$$

, where the $n_e$ and $\mu_e$ are the carrier concentration and mobility of electron, respectively, while the $n_h$ and $\mu_h$ are those of hole, respectively. If we also adopt the high field limit condition, *i. e.*, $H \rightarrow \infty$ which is applicable for the $R_H$ of the polycrystalline LaFeAsO$_{1-x}$H$_x$, the expression is reduced to

$$\lim_{H \to \infty} R_H = \frac{1}{e(n_e - n_h)}. \quad (2)$$

In **Fig. 4(g)**, we plot the $T$ dependence of the effective carrier number per Fe, $N_{eff.}$, calculated from the equation, $N_{eff.} = V/(2eR_H) = N_e - N_h$, where $N_e$ ($N_h$) is the number of electrons (holes) per Fe, and $V$ is the volume of unit cell. In low temperature, the $N_{eff.}$ approaches $0.1 \sim x$ at $T \sim 50$ K, indicating that the 10% hydride substitution for the oxide ion supplies $0.1e^-$ with the Fe.

## Conclusion

We applied the high-pressure synthesis method and carried out a systematic investigation of the factors controlling the crystal growth of the highest $T_c$ superconductor SmFeAsO$_{1-x}$H$_x$, including the kinds of flux, synthesis pressure, and batch compositions of the starting material. The Na$_3$As + 3NaH + As metal hydride flux-growth yielded sub-millimeter-sized crystals that can be readily separated. In comparison with the previous works using NaCl/KCl or NaAs/KAs flux, this mixture flux much efficiently grew the crystal, namely, very rapidly for just a half hour. The chemical composition analyses confirmed 10% substitution of hydrogen for the oxygen site ($x$ = 0.10), however, the structural analyses suggested that the obtained crystal forms a multi-domain structure. By using the FIB technique we fabricated the single domain SmFeAsO$_{0.9}$H$_{0.10}$ crystal with the $T_c$ of 42 K, consistent with 43 K confirmed by the susceptibility measurement. The in-plane resistivity ($\rho^{ab}$) was metallic, while the out-of-plane ($\rho^c$) semiconducting. The larger resistivity anisotropy ($\rho^c/\rho^{ab}$) of 7.8 was obtained at $T$ = 50 K compared with those of the 11 (3-4), 111 (2.6), 122 (4.2) and 245-type (7.5), reflecting the layered structure of 1111-type. The square-root dependence of $\rho^{ab}$ on temperature suggest that the transport properties of SmFeAsO$_{0.9}$H$_{0.10}$ is governed by the spin-frozen state where the fluctuation frequency of spin are very slow compared with that of orbital. Finally, based on the in-plane Hall coefficient, we confirmed that the dominant carrier of SmFeAsO$_{0.9}$H$_{0.10}$ crystal is an electron, and the 10% substitution of hydrogen for the oxygen site effectively supplies 0.1$e^-$ per Fe, which follows the equation: $O^{2-} \rightarrow H^- + e^-$.

## Acknowledgement

This study was supported by the MEXT Element Strategy Initiative Project to form a research core and Grant-in-Aid for Young Scientists (B) (Grant No. 26800182) from JSPS.

## Competing financial interests

The authors declare no competing financial interests.

## References

[1] Y. Kamihara, T. Watanabe, M. Hirano, and H. Hosono, *J. Am. Chem. Soc.* 130, 3296–3297 (2008).

[2] H. Hosono, and K. Kuroki, *Phys. C Supercond. Its Appl.* 514, 399–422 (2015).

[3] H. Hosono, K. Tanabe, E. Takayama-Muromachi, H. Kageyama, S. Yamanaka, H. Kumakura, M. Nohara, H. Hiramatsu, and S. Fujitsu, *Sci. Technol. Adv. Mater.* 16, 033503 (2015).

[4] C. de la Cruz, Q. Huang, J. W. Lynn, J. Li, W. R. Ii, J. L. Zarestky, H. A. Mook, G. F. Chen, J. L. Luo, N. L. Wang, and P. Dai, *Nature* 453, 899–902 (2008).

[5] T. Nomura, S. W. Kim, Y. Kamihara, M. M. Hirano, P. V. Sushko, K. Kato, M. Takata, A. L. Shluger, and H. Hosono, *Supercond. Sci. Technol.* 21, 125028 (2008).

[6] R. Zhi-An, L. Wei, Y. Jie, Y. Wei, S. Xiao-Li, Zheng-Cai, C. Guang-Can, D. Xiao-Li, S. Li-Ling, Zhou Fang, and Z. Zhong-Xian, *Chin. Phys. Lett.* 25, 2215 (2008).

[7] M. Fujioka, S.J. Denholme, T. Ozaki, H. Okazaki, K. Deguchi, S. Demura, H. Hara, T. Watanabe, H. Takeya, T. Yamaguchi, H. Kumakura, and Y. Takano, *Supercond. Sci. Technol.* 26, 085023 (2013).

[8] J. Zhao, Q. Huang, C. de la Cruz, S. Li, J. W. Lynn, Y. Chen, M. A. Green, G. F. Chen, G. Li, Z. Li, J. L. Luo, N. L. Wang, and P. Dai, *Nat. Mater.* 7, 953–959 (2008).

[9] A. Köhler and G. Behr, *J. Supercond. Nov. Magn.* 22, 565–567 (2009).

[10] T. Hanna, Y. Muraba, S. Matsuishi, N. Igawa, K. Kodama, S. Shamoto, and H. Hosono, *Phys. Rev. B* 84, 024521 (2011).

[11] S. Matsuishi, T. Hanna, Y. Muraba, S. W. Kim, J. E. Kim, M. Takata, S. Shamoto, R. I. Smith, and H. Hosono, *Phys. Rev. B* 85, 014514 (2012).

[12] S. Iimura, S. Matsuishi, and H. Hosono, *Phys. Rev. B* 94, 024512 (2016).

[13] S. Iimura, S. Matuishi, H. Sato, T. Hanna, Y. Muraba, S. W. Kim, J. E. Kim, M. Takata, and H. Hosono, *Nat. Commun.* 3, 943 (2012).

[14] M. Hiraishi, S. Iimura, K. M. Kojima, J. Yamaura, H. Hiraka, K. Ikeda, P. Miao, Y. Ishikawa, S. Torii, M. Miyazaki, I. Yamauchi, A. Koda, K. Ishii, M. Yoshida, J. Mizuki, R. Kadono, R. Kumai, T. Kamiyama, T. Otomo, Y. Murakami, S. Matsuishi, and H. Hosono, *Nat. Phys.* 10, 300–303 (2014).

[15] S. Iimura, H. Okanishi, S. Matsuishi, T. Honda, K. Ikeda, Thomas C. Hansen, T. Otomo, H. Hosono *Proc. Natl. Acad. Sci.* 201703295 (2017). doi:10.1073/pnas.1703295114

[16] Zhigadlo, N. D. *et al.* Single crystals of superconducting SmFeAsO 1− x F y grown at high pressure. *J. Phys. Condens. Matter* **20,** 342202 (2008).

[17] J. Karpinski, N. D. Zhigadlo, S. Katrych, Z. Bukowski, P. Moll, S. Weyeneth, H. Keller, R. Puzniak, M. Tortello, D. Daghero, R. Gonnelli, I. Maggio-Aprile, Y. Fasano, Ø. Fischer, K. Rogacki, and B. Batlogg, *Phys. C Supercond.* 469, 370–380 (2009).

[18] N. D. Zhigadlo, S. Weyeneth, S. Katrych, P. J. W. Moll, K. Rogacki, S. Bosma, R. Puzniak, J. Karpinski, and B. Batlogg, B. *Phys. Rev. B* 86, 214509 (2012).

[19] A. Pisoni, S. Katrych, A. Arakcheeva, T. Verebélyi, M. Bokor, P. Huang, R. Gaál, P. Matus, J. Karpinski, and L. Forró, *Phys. Rev. B* 94, 024525 (2016)..

[20] Y. Fukai and N. Ōkuma, *Jpn. J. Appl. Phys.* 32, L1256 (1993).

[21] Y. Fukai and N. Ōkuma, *Phys. Rev. Lett.* 73, 1640–1643 (1994).

[22] G. Andersson, B. Sundqvist, and G. Backstrom, *J. Appl. Phys.* 65, 3943–3950 (1989).

[23] P. J. W. Moll, R. Puzniak, F. Balakirev, K. Rogacki, J. Karpinski, N. D. Zhigadlo, and B. Batlogg, *Nat. Mater.* 9, 628–633 (2010).

[24] S. I. Vedeneev, B. A. Piot, D. K. Maude, and A. V. Sadakov, *Phys. Rev. B* 87, 134512 (2013).

[25] K. Cho, H. Kim, M. A. Tanatar, Y. J. Song, Y. S. Kwon, W. A. Coniglio, C. C. Agosta, A. Gurevich, and R. Prozorov, *Phys. Rev. B* 83, 060502 (2011).

[26] M. A. Tanatar, N. Ni, C. Martin, R. T. Gordon, H. Kim, V. G. Kogan, G. D. Samolyuk, S. L. Bud'ko, P. C. Canfield, and R. Prozorov, *Phys. Rev. B* 79, 094507 (2009).

[27] S. Drotziger, P. Schweiss, K. Grube, T. Wolf, P. Adelmann, C. Meingast, and H. Lohneysen, *J. Phys. Soc. Jpn.* 79, 124705 (2010).

[28] H. Lei and C. Petrovic, *Phys. Rev. B* 83, 184504 (2011).

[29] H. D. Wang, C. H. Dong, Z. J. Li, Q. H. Mao, S. S. Zhu, C. M. Feng, H. Q. Yuan, and Ming-Hu Fang, *EPL Europhys. Lett.* 93, 47004 (2011).

[30] X. Dong, K. Jin, D. Yuan, H. Zhou, J. Yuan, Y. Huang, W. Hua, J. Sun, P. Zheng, W. Hu, Y. Mao, M. Ma, G. Zhang, F. Zhou, and Z. Zhao, *Phys. Rev. B* 92, 064515 (2015).

[31] J. Kim, H. Nam, G. Li, A. B. Karki, Z. Wang, Y. Zhu, C. K. Shih, J. Zhang, R. Jin, and E. W. Plummer, *Sci. Rep.* 6, 35365 (2016).

[32] Y. Muraba, S. Matsuishi, and H. Hosono, *Phys. Rev. B* 89, 094501 (2014).

[33] E. H. Brandt, *Phys. Rev. B* 54, 4246–4264 (1996).

[34] M. Imada, A. Fujimori, and Y. Tokura, *Rev. Mod. Phys.* 70, 1039–1263 (1998).

[35] T. Moriya, Y. Takahashi, and K. Ueda, *J. Phys. Soc. Jpn.* 59, 2905–2915 (1990).

[36] Z. P. Yin, K. Haule, and G. Kotliar, *Phys. Rev. B* 86, 195141 (2012).

[37] T. Katsufuji, M. Kasai, and Y. Tokura, *Phys. Rev. Lett.* 76, 126–129 (1996).

[38] C. Mirri, P. Calvani, F. M. Vitucci, A. Perucchi, K. W. Yeh, M. K. Wu, and S. Lupi, *Supercond. Sci. Technol.* 25, 045002 (2012).

[39] N. Stojilovic, A. Koncz, L. W. Kohlman, R. Hu, C. Petrovic, and S. V. Dordevic, *Phys. Rev. B* 81, 174518 (2010).

[40] C. Hurd, "The Hall Effect in Metals and Alloys", Springer Science & Business Media, (2012).

**Legends of Figure and Table**

Fig. 1: (a, b) Crystal structure and phase diagram of SmFeAsO. (c, d) Sample cell assemblies for the single crystal growth using the $Na_3As$ flux and the synthesis of polycrystalline samples, and the crystal growth using the $3NaH + As$ or $Na_3As + 3NaH + As$ flux. Light yellow, black, while, green, blue, and grey regions denote the 90wt%NaCl + 10wt%$ZrO_2$, graphite, BN, sample (+ flux) pellet, mixture of $NaBH_4$ and $Ca(OH)_2$, and Mo discs, respectively.

Fig. 2: Optical microscope images of $SmFeAsO_{1-x}H_x$ single crystals.

Fig. 3: (a) XRD pattern of the crystal grown by GC4. A small diffraction peak at $2\theta \sim 44°$ originates from the sample stage. (b) Out-of-plane rocking curve of 003 diffraction. (c) Pole figure of 012 diffraction. (d) Cross-section of Fig.3(c). (e, f) $x$ dependence of the $a$ and $c$ axis lengths of polycrystalline $SmFeAsO_{1-x}H_x$, where the red solid and blue dashed lines denote the $a$ and $c$ axis lengths estimated by the Pawley method performed on the grounded crystals picked up from the GC4-batch, while the blue dotted line is the $c$-axis length estimated by the Pawley method applied for the $\theta$-$2\theta$ scan profile shown in Fig. 3(a).

Fig. 4: (a, b) Secondary electron microscope images of resistance bars used for the anisotropic resistivity and Hall measurements. (c) Temperature dependence of the anisotropic resistivity and the resistivity ratio. (d) Susceptibility curve with zero-field-

cooling under the magnetic field (H) of 10 Oe. (e) Log $[\rho(T) - \rho_0]$ vs log $T$ plots for data on the normal conducting region, where $\rho_0$ is the residual resistivity obtained from fitting (see text). (f, g) Temperature dependence of in-plane Hall coefficient and effective carrier number per Fe.

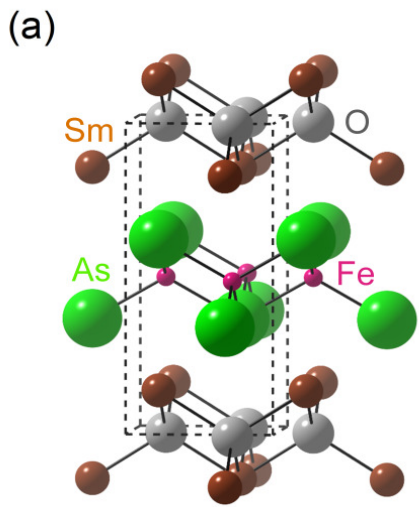
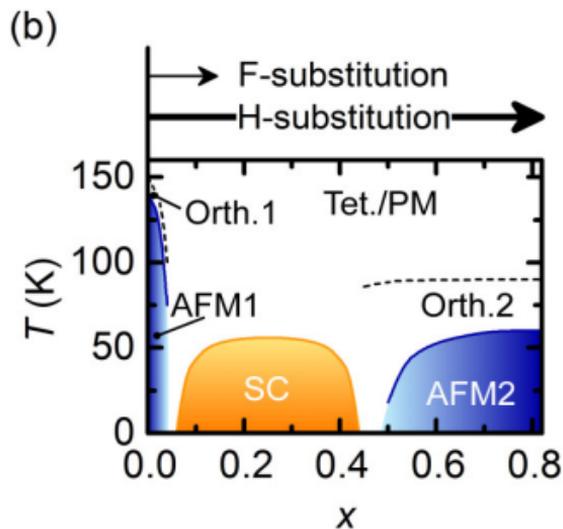
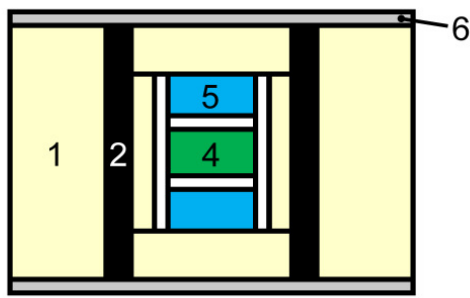
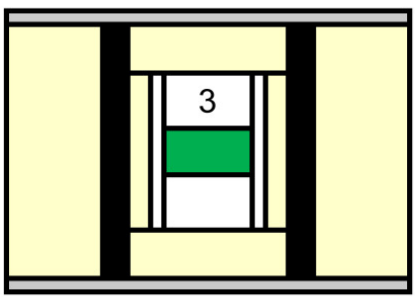

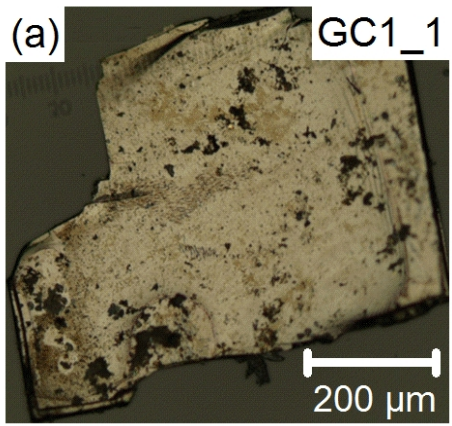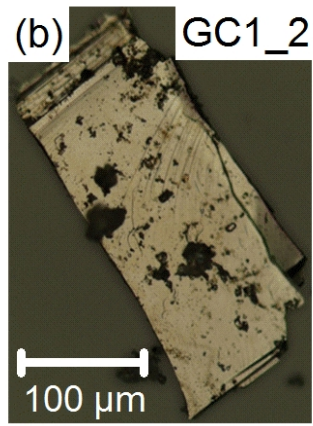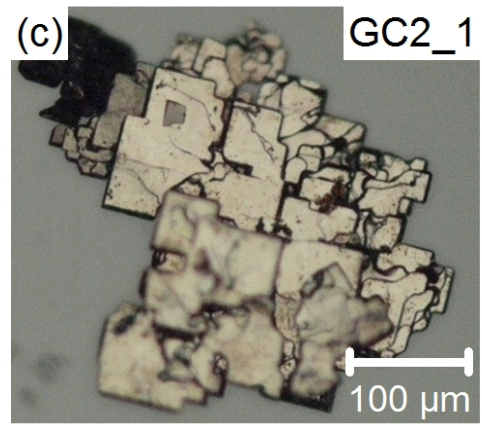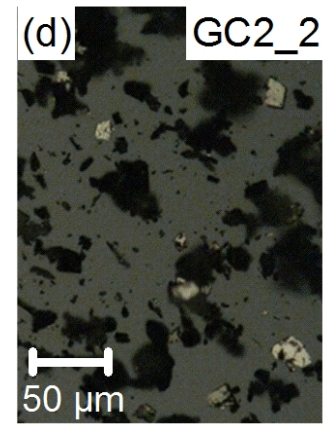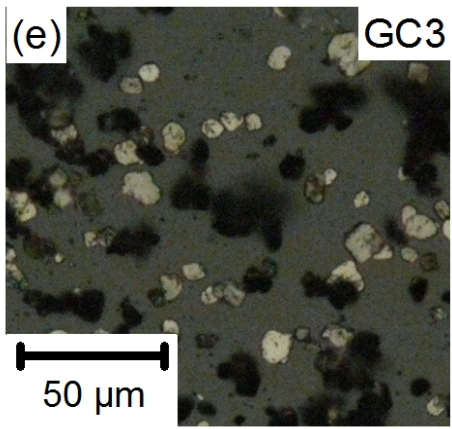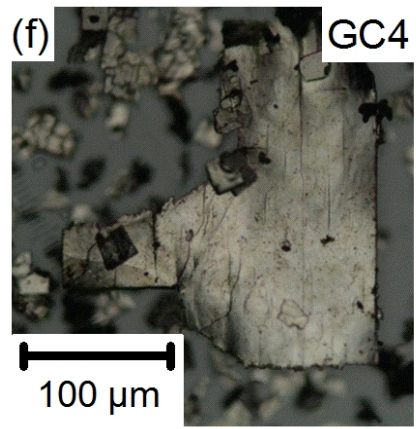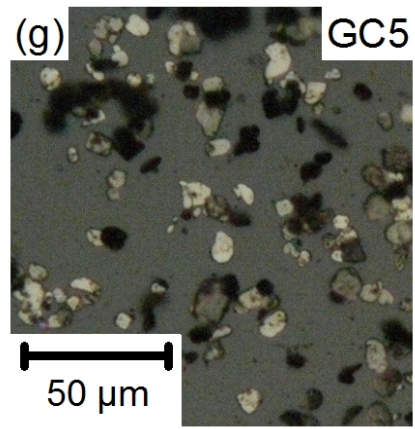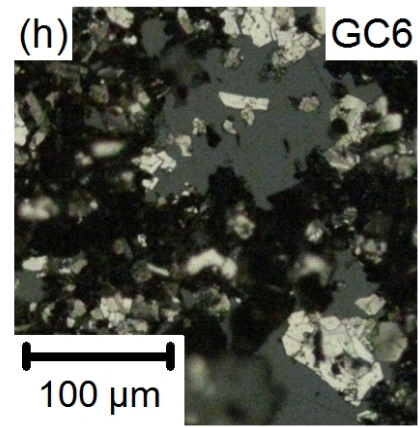

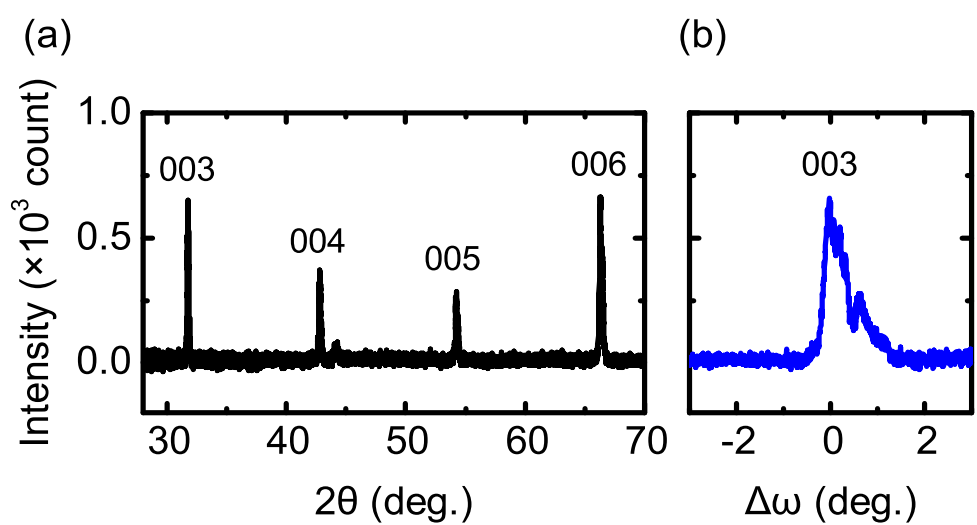
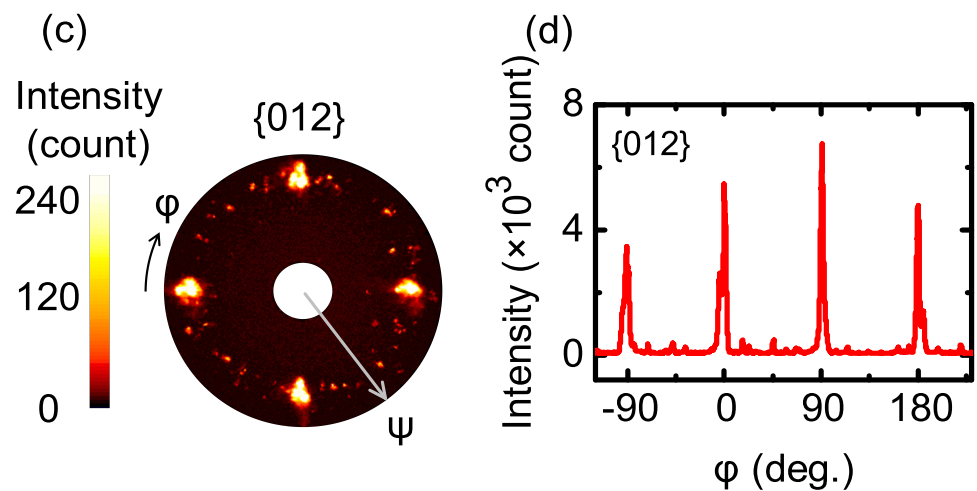
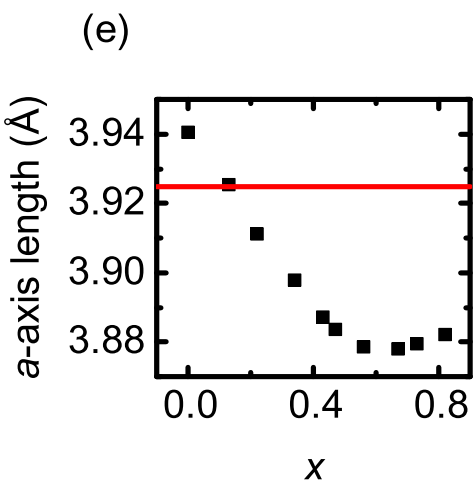
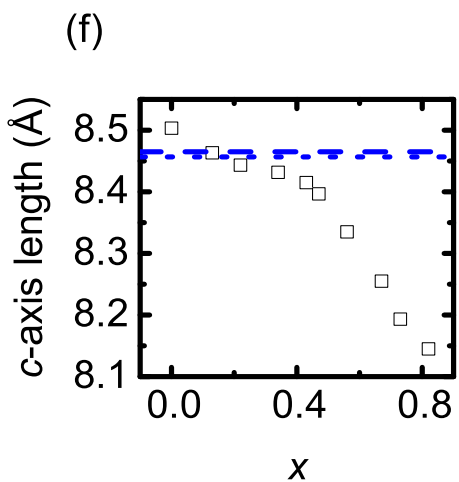

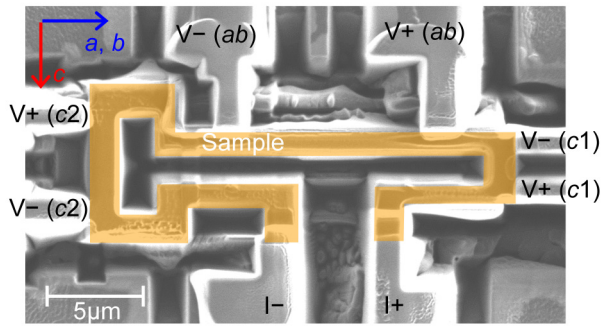
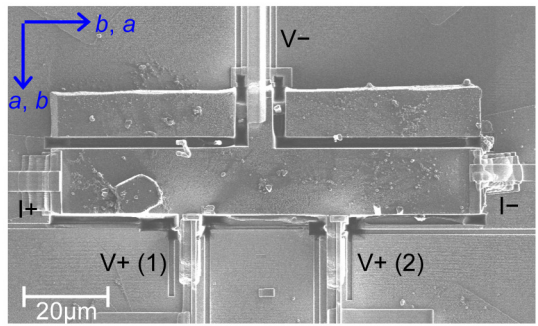
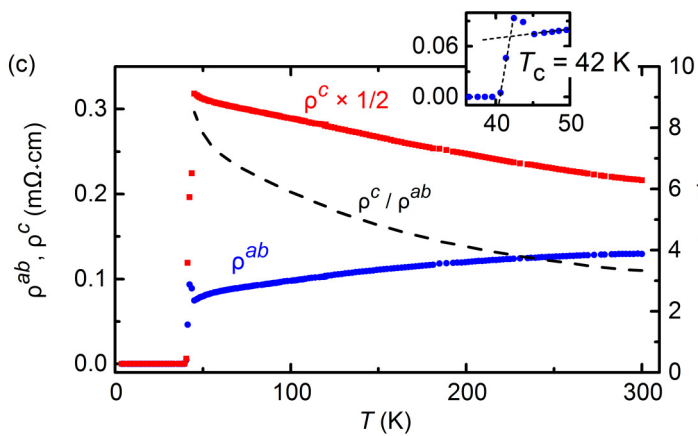
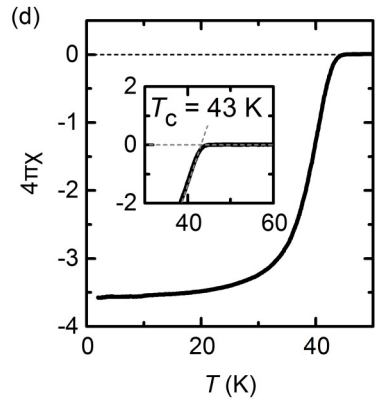
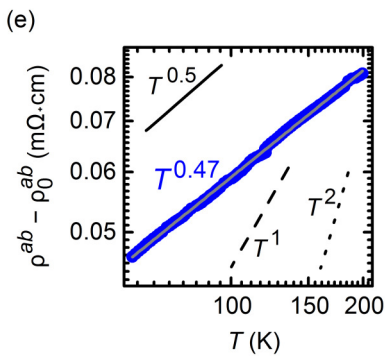
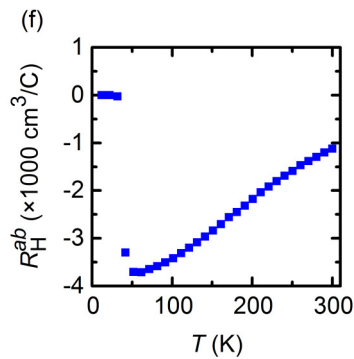
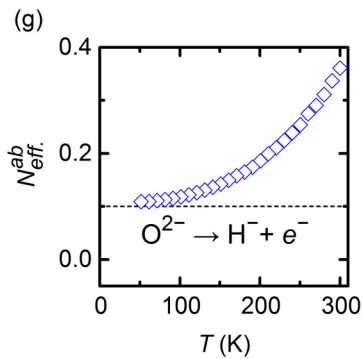

Table 1: Summary of growth conditions (GC) applied in this study and analyzed chemical compositions of resulting SmFeAsO$_{1-x}$H$_x$ crystals. The uncertainty of oxygen contents is a standard deviation calculated from 5-10 data points of oxygen contents separated by micrometer-order in space, while that of hydrogen contents is derived from the uncertainty of balance, TDS repeatability, and integration of TDS-profile with *m/z* =2 to subtract the background which contains hydrogen from adsorbed water on the sample surface.

| Growth Condition | Flux | sample : flux (weight) | *P* / GPa | Nominal Composition | | Analyzed Composition | |
|---|---|---|---|---|---|---|---|
| | | | | O/Sm | H/Sm | O/Sm | H/Sm |
| GC1 | Na$_3$As | 1:5 | 3.0 | 0.6 | 0.4 | 0.90(2) | 0.02(1) |
| GC2 | 3NaH + As | 1:5 | 3.0 | 0.6 | 0.4 | 0.90(2) | 0.09(1) |
| GC3 | 3NaH + As | 1:10 | 3.0 | 0.6 | 0.4 | 0.96(2) | 0.070(2) |
| GC4 | Mixture | 1:10 | 3.0 | 0.6 | 0.4 | 0.89(2) | 0.100(6) |
| GC5 | 3NaH + As | 1:10 | 5.5 | 0.6 | 0.4 | 0.79(2) | 0.163(5) |
| GC6 | Mixture | 1:10 | 5.5 | 0.6 | 0.4 | 0.79(3) | 0.177(3) |
| GC7 | Mixture | 1:10 | 5.5 | 0.8 | 0.2 | 0.87(2) | 0.13(1) |
| GC8 | Mixture | 1:10 | 5.5 | 0.8 | 0.8 | 0.84(2) | 0.155(9) |
| GC9 | Mixture | 1:10 | 5.5 | 0.7 | 1.0 | 0.81(3) | 0.156(2) |

Table2: Resistivity anisotropy ($\rho^c/\rho^{ab}$) of several iron-based superconductors at $T$ = 50 K.

| System | Compound | $\rho^c/\rho^{ab}$ at 50 K | Inter-layer distance (Å) | Reference |
|---|---|---|---|---|
| 11 | FeSe | 3-4 | 5.49 | 24 |
| 111 | LiFeAs | 2.6 | 6.3463 | 25 |
| 122 | Ba(Fe$_{0.926}$Co$_{0.074}$)$_2$As$_2$ | 4.2 | 6.50 | 26,27 |
| 245 | K$_{0.64}$Fe$_{1.44}$Se$_{2.00}$ | 7.5 | 7.0375 | 28 |
|  | K$_{0.58}$Rb$_{0.42}$Fe$_{1.72}$Se$_2$ | 45.7 | 7.1515 | 29 |
| 1111 | SmFeAsO$_{0.9}$H$_{0.10}$ | 7.8 | 8.456 | Present work |
|  | SmFeAsO$_{0.7}$F$_{0.25}$ | 8.4 | 8.4655 | 17,23 |
| 11111 | (Li$_{0.84}$Fe$_{0.16}$)OHFe$_{0.98}$Se | 2500 | 9.3184 | 30 |
| 1048 | Ca$_{10}$Pt$_4$As$_8$(Fe$_{2-x}$Pt$_x$As$_2$)$_5$ | 56 | 11.2 | 31 |